\documentclass[pre,aps,preprint,showpacs,preprintnumbers,amsmath,amssymb,secnumroman,eqsecnum]{revtex4}

\usepackage{graphicx}
\usepackage{dcolumn}
\usepackage{bm}

\newcommand{\beq}{\begin{equation}}
\newcommand{\eeq}{\end{equation}}
\newcommand{\beqa}{\begin{eqnarray}}
\newcommand{\eeqa}{\end{eqnarray}}
\newcommand{\vc}[1]{\mbox{\boldmath $#1$}}

\newcommand{\vol}[1]{{\bf #1}}

\newcommand{\du}[1]{{\bf\sf #1}}

\begin{document}


\title{Stokesian swimming of a prolate spheroid at low Reynolds number}

\author{B. U. Felderhof}

 \email{ufelder@physik.rwth-aachen.de}
\affiliation{Institut f\"ur Theorie der Statistischen Physik\\ RWTH Aachen University\\
Templergraben 55\\52056 Aachen\\ Germany\\
}%



\date{\today}

\begin{abstract}
The swimming of a prolate spheroid immersed in a viscous incompressible fluid and performing surface deformations periodically in time is studied on the basis of Stokes' equations of low Reynolds number hydrodynamics. The average over a period of time of the translational and rotational swimming velocity and the rate of dissipation are given by integral expressions of second order in the amplitude of surface deformations. The first order flow velocity and pressure, as functions of prolate spheroidal coordinates, are expressed as sums of basic solutions of Stokes' equations. Sets of superposition coefficients of these solutions which optimize the mean translational swimming speed for given power are derived from an eigenvalue problem. The maximum eigenvalue is a measure of the efficiency of the  optimal stroke within the chosen class of motions. The maximum eigenvalue for sets of low multipole order is found to be a strongly increasing function of the aspect ratio of the spheroid.
\end{abstract}

\maketitle

\section{\label{I}Introduction}

The theory of the swimming of micro-organisms in a viscous incompressible fluid is based on the Stokes equations of low Reynolds number hydrodynamics \cite{1}. In this regime inertia plays no role and the effect of swimming can be understood on the basis of purely viscous flow. It was shown by Taylor \cite{2} in the example of the swimming of a planar sheet distorted by a transverse surface wave that the effect is purely kinematic. His calculation to second order in the wave amplitude leads to a swimming velocity which is independent of the viscosity of the fluid. Taylor's analysis was subsequently extended to a squirming sphere by Lighthill \cite{3}. Blake corrected this work and applied it in a spherical envelope approach to ciliary propulsion \cite{4}.

The calculations of mean translational swimming velocity and rate of dissipation to second order in the amplitude of surface distortion are complicated, and for simplicity, the analysis for a sphere was restricted to axial strokes. The extension to general strokes was derived only recently \cite{5}. The extension also allows the mean rotational swimming velocity achieved by a general stroke to be calculated.

In the following we consider Stokesian swimming of a prolate spheroid, again to second order in the amplitude of surface distortion. Since arbitrary aspect ratio is allowed, the model provides interesting physical applications. In particular the swimming of Paramecium should be well described by the model. Paramecium rotates about its long axis as it swims, so that it is important to consider both the mean translational and rotational swimming velocity. The motion of a prolate spheroid on the basis of active particle theory was studied by Leshansky et al. \cite{6} for a particular mode of steady tangential surface distortion.

The reciprocal theorem \cite{1} is used to derive integral expressions for the mean translational and rotational swimming velocity which are bilinear in the amplitude of surface displacements. We show how these expressions, when written in terms of prolate spheroidal coordinates, can be reduced to one-dimensional integrals over the polar variable. In the process we encounter identities which have been known for a long time in terms of ellipsoidal coordinates \cite{7},\cite{8},\cite{9}. The validity of the identities was explained recently by Kim \cite{10},\cite{11} on the basis of a symmetry property of the Stokes double-layer operator \cite{12}.

In the calculation, the first order flow velocities and pressures are expanded in terms of a basic set of solutions of the steady state Stokes equations. We derive these solutions as functions of prolate spheroidal coordinates. The mean swimming velocities and the mean rate of dissipation become bilinear expressions in terms of the amplitudes of the mode functions involving three matrices corresponding to the chosen representation. As usual, optimization of the translational swimming speed for given power leads to an eigenvalue problem \cite{13}-\cite{17}. The maximum eigenvalue is a measure of the efficiency of the optimal stroke within the chosen class of motions. In contrast to the case of a sphere, the three matrices cannot be evaluated analytically. We derive numerical results for the maximum eigenvalue for a wide range of aspect ratios, as well as for the corresponding mean rate of rotation.

\section{\label{II}Swimming velocity and power}

We consider a prolate spheroid of major semi-axis $a$ and minor semi-axis $b$ immersed in a viscous
incompressible fluid of shear viscosity $\eta_s$. We choose Cartesian coordinates such that the $z$ axis is in the direction of the long axis. At low Reynolds
number and on a slow time scale the flow velocity
$\vc{v}(\vc{r},t)$ and the pressure $p(\vc{r},t)$ satisfy the
Stokes equations
\begin{equation}
\label{2.1}\eta_s\nabla^2\vc{v}-\nabla p=0,\qquad\nabla\cdot\vc{v}=0.
\end{equation}
The fluid is set in motion by distortions of the
surface which are periodic in time and lead to a time-dependent flow field as well as to a swimming
motion of the spheroid. The surface displacement
$\vc{\xi}(\vc{s},t)$ is defined as the vector distance
\begin{equation}
\label{2.2}\vc{\xi}=\vc{s}'-\vc{s}
\end{equation}
of a point $\vc{s}'$ on the displaced surface $S(t)$ from the
point $\vc{s}$ on the spheroid with surface $S_0$. The fluid
velocity $\vc{v}(\vc{r},t)$ in the rest frame is required to satisfy \cite{15}
\begin{equation}
\label{2.3}\vc{v}(\vc{s}+\vc{\xi}(\vc{s},t))=\frac{\partial\vc{\xi}(\vc{s},t)}{\partial t},
\end{equation}
corresponding to a no-slip boundary condition. The instantaneous translational swimming velocity $\vc{U}(t)$,
the rotational swimming velocity $\vc{\Omega}(t)$, and the flow pattern $(\vc{v},p)$ follow from the condition that no net
force or torque is exerted on the fluid. We evaluate these quantities by a perturbation expansion in powers of the
displacement $\vc{\xi}(\vc{s},t)$.

To second order in $\vc{\xi}$ the flow velocity and the swimming velocity
take the form \cite{15}
\begin{equation}
\label{2.4}\vc{v}(\vc{r},t)=\vc{v}_1(\vc{r},t)+\vc{v}_2(\vc{r},t)+...,\qquad
\vc{U}(t)=\vc{U}_2(t)+....
\end{equation}
Both $\vc{v}_1$ and $\vc{\xi}$ are assumed to vary harmonically with frequency
$\omega$, and can be expressed as
 \begin{eqnarray}
\label{2.5}\vc{v}_1(\vc{r},t)&=&\vc{v}_{1c}(\vc{r})\cos\omega t+\vc{v}_{1s}(\vc{r})\sin\omega t,\nonumber\\
\vc{\xi}(\vc{s},t)&=&\vc{\xi}_{c}(\vc{s})\cos\omega
t+\vc{\xi}_{s}(\vc{s})\sin\omega t.
\end{eqnarray}
Expanding the no-slip condition Eq. (2.3) to second order we find
for the flow velocity at the surface
\begin{eqnarray}
\label{2.6}\vc{u}_{1S}(\vc{s},t)&=&\vc{v}_1\big|_{S_0}=\frac{\partial\vc{\xi}(\vc{s},t)}{\partial t},\nonumber\\
\vc{u}_{2S}(\vc{s},t)&=&\vc{v}_2\big|_{S_0}=-\vc{\xi}\cdot\nabla\vc{v}_1\big|_{S_0}.
\end{eqnarray}
In complex notation with $\vc{v}_1=\vc{v}_\omega\exp(-i\omega t)$ the mean second order surface velocity is given by
\begin{equation}
\label{2.7}\overline{\vc{u}}_{2S}(\vc{s})=-\frac{1}{2}\mathrm{Re}(\vc{\xi}^*_\omega\cdot\nabla)\vc{v}_\omega\big|_{S_0},
\end{equation}
where the overhead bar indicates a time-average over a period $T=2\pi/\omega$.

The time-averaged second order flow velocity $\overline{\vc{v}^{(2)}}(\vc{r})$ and corresponding mean pressure $\overline{p^{(2)}}(\vc{r})$ satisfy the Stokes equations Eq. (2.1) with boundary value $\overline{\vc{v}^{(2)}}(\vc{s})=\overline{\vc{u}}_{2S}(\vc{s})$. Moreover the flow tends to $-\overline{\vc{U}^{(2)}}-\overline{\vc{\Omega}^{(2)}}\times\vc{r}$ at infinity and satisfies the condition of vanishing hydrodynamic force and torque. In the laboratory frame this corresponds to the flow $(\overline{\vc{u}^{(2)}}(\vc{r}),\overline{p^{(2)}}(\vc{r}))$ with
\begin{equation}
\label{2.8}\overline{\vc{u}^{(2)}}(\vc{r})=\overline{\vc{U}^{(2)}}+\overline{\vc{\Omega}^{(2)}}\times\vc{r}+\overline{\vc{v}^{(2)}}(\vc{r}).
\end{equation}
As a second flow we consider the solution of the Stokes friction problem for solid body motion of the surface $S_0$ with force $\vc{\mathcal{F}}$ and torque $\vc{\mathcal{T}}$ exerted on the fluid. Applying the reciprocal theorem \cite{1} to the pair of flows we find the relation
\begin{equation}
\label{2.9}\vc{\mathcal{F}}\cdot\overline{\vc{U}^{(2)}}+\vc{\mathcal{T}}\cdot\overline{\vc{\Omega}^{(2)}}
=\int_{S_0}\vc{n}\cdot\vc{\sigma}_{fr}\cdot\overline{\vc{u}}_{2S}\;dS_0,
\end{equation}
where $\vc{n}$ is the outward normal to the surface $S_0$ and $\vc{\sigma}_{fr}$ is the stress tensor for the Stokes friction problem with translation and rotation.

We consider periodic surface distortions such that the mean translational displacement of the spheroid is in the $z$ direction and the mean rotation is about the $z$ axis. For the second order mean translational swimming velocity $\overline{U}_2$ we have
\begin{equation}
\label{2.10}\overline{U}_2=\frac{1}{\mathcal{F}_z}\int_{S_0}\vc{n}\cdot\vc{\sigma}_{Tz}(\vc{s})\cdot\overline{\vc{u}}_{2S}(\vc{s})\;dS_0,
\end{equation}
where $\vc{\sigma}_{Tz}(\vc{s})$ is the stress exerted at the surface $S_0$ when the spheroid with no-slip boundary condition is subjected to a force $\mathcal{F}_z\vc{e}_z$ in the $z$ direction. Similarly the second order mean rotational swimming velocity $\overline{\Omega}_2$ is given by
\begin{equation}
\label{2.11}\overline{\Omega}_2=\frac{1}{\mathcal{T}_z}\int_{S_0}\vc{n}\cdot\vc{\sigma}_{Rz}(\vc{s})\cdot\overline{\vc{u}}_{2S}(\vc{s})\;dS_0,
\end{equation}
where $\vc{\sigma}_{Rz}(\vc{s})$ is the stress exerted at the surface $S_0$ when the spheroid with no-slip boundary condition is subjected to a torque $\mathcal{T}_z\vc{e}_z$ in the $z$ direction.

To second order the mean rate of dissipation $\overline{\mathcal{D}}_2$ is
determined entirely by the first order solution. It may be
expressed as a surface integral \cite{15}
\begin{equation}
\label{2.12}\overline{\mathcal{D}}_2=-\frac{1}{2}\mathrm{Re}\int_{S_0}\vc{v}^*_\omega\cdot\vc{\sigma}_\omega\cdot\vc{n}\;dS_0,
\end{equation}
where $\vc{\sigma}_\omega$ is the first order stress tensor, given by
\begin{equation}
\label{2.13}\vc{\sigma}_\omega=\eta_s(\nabla\vc{v}_\omega+[\nabla\vc{v}_\omega]^T)-p_\omega\vc{I}.
\end{equation}
The mean rate of dissipation equals the power necessary to generate the motion.

\section{\label{III}Mode functions}

We use spheroidal coordinates $(\xi,\eta,\varphi)$ in which the Cartesian coordinates $(x,y,z)$ are expressed as
\begin{eqnarray}
\label{3.1}x&=&c\sqrt{(\xi^2-1)(1-\eta^2)}\cos\varphi,\nonumber\\
y&=&c\sqrt{(\xi^2-1)(1-\eta^2)}\sin\varphi,\nonumber\\
z&=&-c\xi\eta,
\end{eqnarray}
where $c=\sqrt{a^2-b^2}$ is the semi-focal distance. The surface of the spheroid corresponds to the value $\xi_0$ given by
\begin{equation}
\label{3.2}a=c\xi_0,\qquad b=c\sqrt{\xi_0^2-1}.
\end{equation}
The coordinates vary in the ranges $\xi_0<\xi<\infty,\;-1<\eta<1,\;0<\varphi<2\pi$. For large $\xi$ the surface $\xi=constant$ becomes spherical and the variable $\eta$ can be identified with $-\cos\theta$, where $\theta$ is the polar angle. The variable $\varphi$ is the azimuthal angle. The coordinates $(\xi,\eta,\varphi)$ are identical to those introduced by Morse and Feshbach \cite{18}, except for the minus sign in the last line of Eq. (3.1). Our choice guarantees a right-handed system.

The metric coefficients are given by
\begin{equation}
\label{3.3}h_1=\frac{1}{c}\sqrt{\frac{\xi^2-1}{\xi^2-\eta^2}},\qquad h_2=\frac{1}{c}\sqrt{\frac{1-\eta^2}{\xi^2-\eta^2}},\qquad h_3=\frac{1}{c}\sqrt{\frac{1}{(\xi^2-1)(1-\eta^2)}}.
\end{equation}
These can be used to express the required differential operators \cite{1}.

We consider functions $\vc{v}_\omega(\vc{r})$ with dependence on $\varphi$ given by a factor $\exp(im\varphi)$. Then the integrals in Eqs. (2.8)-(2.10) can be reduced to integrals over the variable $\eta$. A set of basic solutions of the Stokes equations can be derived from scalar functions of the factorized form
\begin{equation}
\label{3.4}\Phi^m_n(\vc{r})=Q^m_n(\xi)P^m_n(\eta)e^{im\varphi},
\end{equation}
where $Q^m_n(\xi)$ is an associated Legendre function of the second kind, and $P^m_n(\eta)$ is an associated Legendre function of the first kind in the notation of Edmonds \cite{19}. The function $Q^m_n(\xi)$ is defined such that it tends to zero at infinity. It can be expressed as \cite{20}
\begin{equation}
\label{3.5}Q^m_n(\xi)=(\xi^2-1)^{m/2}\frac{d^m}{d\xi^m}\;Q_n(\xi)
\end{equation}
with
\begin{equation}
\label{3.6}Q_n(\xi)=\int^\infty_0[\xi+\sqrt{\xi^2-1}\cosh\alpha]^{-n-1}\;d\alpha,\qquad \xi>1.
\end{equation}
For large $\xi$,
\begin{equation}
\label{3.7}Q^m_n(\xi)=(-1)^m\frac{\sqrt{\pi}\;\Gamma(n+m+1)}{\Gamma(n+3/2)}\;(2\xi)^{-n-1}\bigg[1+O\bigg(\frac{1}{\xi}\bigg)\bigg].
\end{equation}
The function $\Phi^m_n(\vc{r})$ in Eq. (3.4) satisfies Laplace's equation.

We derive a corresponding set of solutions of the Stokes equations Eq. (2.1) with vanishing pressure disturbance as
\begin{equation}
\label{3.8}\vc{u}_{nm}(\vc{r})=c\nabla\Phi^m_n(\vc{r}).
\end{equation}
A second independent set of solutions is defined by
\begin{equation}
\label{3.9}\vc{w}_{nm}(\vc{r})=-i\frac{\vc{r}}{c}\times\vc{u}_{nm}(\vc{r}).
\end{equation}
These functions also satisfy Eq. (2.1) with vanishing pressure disturbance. We have chosen a phase factor in accordance with the choice made for flow about a sphere \cite{5}. A third set of solutions of Eq. (2.1) with nonvanishing pressure disturbance can be constructed from the potentials in Eq. (3.4) by the device of Papkovich \cite{21} and Neuber \cite{22},\cite{23}. These flows take the form
\begin{equation}
\label{3.10}\vc{v}_{nm}(\vc{r})=z\nabla\Phi^m_{n-1}(\vc{r})-\vc{e}_z\Phi^m_{n-1}(\vc{r}),
\end{equation}
where $\vc{e}_z$ is the unit vector in the $z$ direction. The corresponding pressure disturbance is given by
\begin{equation}
\label{3.11}p_{nm}(\vc{r})=2\eta_s\vc{e}_z\cdot\nabla\Phi^m_{n-1}(\vc{r}).
\end{equation}
The shift in the index $n$ is useful for the comparison with flow about a sphere \cite{24},\cite{25}. In Eq. (3.10) the index $m$ takes integer values $-n+1,...,n-1$, so that for each $n$ there are $2n-1$ vector functions and corresponding scalars in Eq. (3.11). We define two more vector functions for each $n$ as
\begin{equation}
\label{3.12}\vc{v}_{n,\pm n}(\vc{r})=(x\pm i y)\nabla\Phi^{\pm n\mp 1}_{n-1}(\vc{r})-(\vc{e}_x\pm i \vc{e}_y)\Phi^{\pm n\mp 1}_{n-1}(\vc{r}),
\end{equation}
with the corresponding pressure functions
\begin{equation}
\label{3.13}p_{n,\pm n}(\vc{r})=2\eta_s(\vc{e}_x\pm i\vc{e}_y)\cdot\nabla\Phi^{\pm n\mp 1}_{n-1}(\vc{r}).
\end{equation}

The general solution of Eq. (2.1) which tends to zero at infinity and varies harmonically in time can be expressed as the complex flow velocity and pressure
\begin{eqnarray}
\label{3.14}\vc{v}^c_1(\vc{r},t)&=&-\omega a\sum^\infty_{n=1}\sum^n_{m=-n}\bigg[\kappa_{nm}\vc{v}_{nm}(\vc{r})+\nu_{nm}\vc{w}_{nm}(\vc{r})+\mu_{nm}\vc{u}_{nm}(\vc{r})\bigg]e^{-i\omega t},\nonumber\\
p^c_1(\vc{r},t)&=&-\omega a\sum^\infty_{n=1}\sum^{n}_{m=-n}\kappa_{nm}p_{nm}(\vc{r})e^{-i\omega t},
\end{eqnarray}
with complex coefficients $(\kappa_{nm},\nu_{nm},\mu_{nm})$. The solutions contain a factor $\exp[i(m\varphi-\omega t)]$, representing a running wave in the azimuthal direction for $m\neq0$.

\section{\label{IV}Mean translational swimming velocity}

In order to calculate the mean translational swimming velocity we must find the vector function $\vc{n}\cdot\vc{\sigma}_{Tz}(\vc{s})$ occurring in Eq. (2.10). To this purpose we must solve the Stokes friction problem for a force $\mathcal{F}_z\vc{e}_z$ acting on the spheroid in the $z$ direction. We define a dimensionless stress tensor $\hat{\vc{\sigma}}_{Tz}$ by putting
\begin{equation}
\label{4.1}\vc{\sigma}_{Tz}=U\frac{\eta_s}{c}\;\hat{\vc{\sigma}}_{Tz},
\end{equation}
where $U$ is the velocity of the Stokes friction problem for force $\mathcal{F}_z$. The friction coefficient $\zeta_T$ is defined by
\begin{equation}
\label{4.2}\mathcal{F}_z=\zeta_TU,\qquad\zeta_T=4\pi\eta_scf_T,
\end{equation}
where $f_T$ is the dimensionless form. With these definitions Eq. (2.10) takes the form
\begin{equation}
\label{4.3}\overline{U}_2=\frac{1}{4\pi f_T}\int\frac{\vc{e}_1\cdot\hat{\vc{\sigma}}_{Tz}\cdot\overline{\vc{u}}_{2S}}{c^2h_2h_3}\bigg|_{\xi=\xi_0}\;d\eta d\varphi,
\end{equation}
since in spheroidal coordinates $\vc{n}=\vc{e}_1=(1,0,0)$ and $dS_0=d\eta d\varphi/(h_2h_3)$.

We solve the Stokes friction problem by means of the Ansatz
\begin{equation}
\label{4.4}\hat{\vc{v}}_{Tz}(\vc{r})=-\vc{e}_z+A_T\hat{\vc{v}}_{10}(\vc{r})+B_T\vc{u}_{10}(\vc{r}),\qquad\hat{p}_{Tz}(\vc{r})=A_Tp_{10}(\vc{r}),
\end{equation}
with coefficients $A_T,B_T$ and the definition
\begin{equation}
\label{4.5}\hat{\vc{v}}_{10}(\vc{r})=\vc{v}_{10}(\vc{r})-\vc{u}_{10}(\vc{r}).
\end{equation}
This linear combination takes the simple form
\begin{equation}
\label{4.6}\hat{\vc{v}}_{10}(\vc{r})=\bigg(\frac{2\eta\xi}{\sqrt{(\xi^2-1)(\xi^2-\eta^2)}},\sqrt{\frac{1-\eta^2}{\xi^2-\eta^2}},0\bigg).
\end{equation}
This decays at large distance as $1/\xi$, whereas the function $\vc{u}_{10}(\vc{r})$ decays as $1/\xi^3$. The coefficient of the $1/\xi$ term in Eq. (4.4) is determined by the friction coefficient with the relation
\begin{equation}
\label{4.7}A_T\hat{\vc{v}}_{10}(\vc{r})=\zeta_T\vc{T}(\vc{r})\cdot\vc{e}_z+O(1/\xi^2),
\end{equation}
where $\vc{T}(\vc{r})$ is the Oseen tensor given by
\begin{equation}
\label{4.8}\vc{T}(\vc{r})=\frac{1}{8\pi\eta_s}\frac{\vc{I}+\hat{\vc{r}}\hat{\vc{r}}}{r}.
\end{equation}
We note that the unit vector $\vc{e}_z$ is given by
\begin{equation}
\label{4.9}\vc{e}_z=\bigg(-\eta\sqrt{\frac{\xi^2-1}{\xi^2-\eta^2}},-\xi\sqrt{\frac{1-\eta^2}{\xi^2-\eta^2}},0\bigg),
\end{equation}
which behaves as $\vc{e}_z\approx (-\eta,-\sqrt{1-\eta^2},0)$ for large $\xi$. From Eq. (4.7) the dimensionless friction coefficient is given by
\begin{equation}
\label{4.10}f_T=-2A_T.
\end{equation}

Applying the no-slip boundary condition at $\xi=\xi_0$ to the velocity field in Eq. (4.4) we obtain the coefficients $A_T,B_T$ as
\begin{equation}
\label{4.11}A_T=\frac{-1}{(\xi_0^2+1)Q_0(\xi_0)-\xi_0},\qquad B_T=\frac{-(\xi_0^2+1)}{(\xi_0^2+1)Q_0(\xi_0)-\xi_0}.
\end{equation}
Using this we find from the velocity field and pressure
\begin{equation}
\label{4.12}\frac{1}{f_Tc^2h_2h_3}\;\vc{e}_1\cdot\hat{\vc{\sigma}}_{Tz}\bigg|_{\xi=\xi_0}=-\vc{e}_z.
\end{equation}
Substituting into Eq. (4.3) we obtain
\begin{equation}
\label{4.13}\overline{U}_2=-\frac{1}{2}\int^1_{-1}\vc{e}_z\cdot\overline{\vc{u}}_{2S}\bigg|_{\xi=\xi_0}\;d\eta,
\end{equation}
for surface velocity $\overline{\vc{u}}_{2S}$ which does not depend on the variable $\varphi$. This reduces to the result for a sphere \cite{16} in the limit $b=a$. For tangential surface velocity the expression reduces to  that derived by Leshansky et al. \cite{6}.

\section{\label{V}Mean rotational swimming velocity}

In order to calculate the mean rotational swimming velocity we must find the vector function $\vc{n}\cdot\vc{\sigma}_{Rz}(\vc{s})$ occurring in Eq. (2.11). To this purpose we must solve the Stokes friction problem for a torque $\mathcal{T}_z\vc{e}_z$ acting on the spheroid in the $z$ direction. We define a dimensionless stress tensor $\hat{\vc{\sigma}}_{Rz}$ by putting
\begin{equation}
\label{5.1}\vc{\sigma}_{Rz}=\Omega\eta_s\hat{\vc{\sigma}}_{Rz},
\end{equation}
where $\Omega$ is the rotational velocity of the Stokes friction problem. The friction coefficient $\zeta_R$ is defined by
\begin{equation}
\label{5.2}\mathcal{T}_z=\zeta_R\Omega,\qquad\zeta_R=4\pi\eta_sc^3f_R,
\end{equation}
where $f_R$ is the dimensionless form. With these definitions Eq. (2.11) takes the form
\begin{equation}
\label{5.3}\overline{\Omega}_2=\frac{1}{4\pi cf_R}\int\frac{\vc{e}_1\cdot\hat{\vc{\sigma}}_{Rz}\cdot\overline{\vc{u}}_{2S}}{c^2h_2h_3}\bigg|_{\xi=\xi_0}\;d\eta d\varphi.
\end{equation}

We solve the Stokes friction problem by means of the Ansatz
\begin{equation}
\label{5.4}\hat{\vc{v}}_{Rz}(\vc{r})=-\vc{e}_z\times\vc{r}+A_R\vc{w}_1(\vc{r})+B_R\vc{w}_2(\vc{r}),
\end{equation}
with coefficients $A_R,B_R$, a first azimuthal flow
\begin{eqnarray}
\label{5.5}\vc{w}_1(\vc{r})&=&i\vc{w}_{10}(\vc{r})\nonumber\\
&=&\bigg(0,0,\frac{1}{\xi^2-\eta^2}\sqrt{\frac{1-\eta^2}{\xi^2-1}}\big[\xi+\eta^2\xi-\xi^3+(\xi^2-\eta^2)(\xi^2-1)Q_0(\xi)\big]\bigg),
\end{eqnarray}
and a second azimuthal flow
\begin{equation}
\label{5.6}\vc{w}_2(\vc{r})=\vc{e}_z\times\vc{u}_{00}(\vc{r})=\bigg(0,0,\frac{-\xi}{\xi^2-\eta^2}\sqrt{\frac{1-\eta^2}{\xi^2-1}}\bigg).
\end{equation}
Here we have used
\begin{equation}
\label{5.7}\vc{u}_{00}(\vc{r})=\bigg(\frac{1}{\sqrt{(\xi^2-1)(\xi^2-\eta^2)}},0,0\bigg).
\end{equation}
At large distance from the origin
\begin{equation}
\label{5.8}\vc{w}_1(\vc{r})\approx\frac{1}{3\xi^2}\sqrt{1-\eta^2}\;\vc{e}_
\varphi,\qquad\vc{w}_2(\vc{r})\approx\frac{-1}{\xi^2}\sqrt{1-\eta^2}\;\vc{e}_
\varphi,
\end{equation}
so that
\begin{equation}
\label{5.9}\hat{\vc{v}}_{Rz}(\vc{r})\approx-\vc{e}_z\times\vc{r}+\bigg(\frac{1}{3}A_R-B_R\bigg)c^2\frac{\vc{e}_z\times\vc{r}}{r^3}.
\end{equation}
Hence we can identify
\begin{equation}
\label{5.10}\frac{1}{3}A_R-B_R=\frac{c}{2}f_R.
\end{equation}

Applying the no-slip boundary condition at $\xi=\xi_0$ to the velocity field in Eq. (5.4) we obtain the coefficients $A_R,B_R$ as
\begin{equation}
\label{5.11}A_R=B_R=c\frac{\xi_0^2-1}{(\xi_0^2-1)Q_0(\xi_0)-\xi_0}.
\end{equation}
This yields for the dimensionless friction coefficient for rotation about the long axis
\begin{equation}
\label{5.12}f_R=\frac{4}{3}\frac{\xi_0^2-1}{\xi_0-(\xi_0^2-1)Q_0(\xi_0)},
\end{equation}
in agreement with the result of Jeffery \cite{26}. The pressure disturbance vanishes.
From the velocity field we find
\begin{equation}
\label{5.13}\frac{1}{f_Rc^2h_2h_3}\;\vc{e}_1\cdot\hat{\vc{\sigma}}_{Rz}\bigg|_{\xi=\xi_0}=-\frac{3}{2c(\xi_0^2-1)}\;\vc{e}_z\times\vc{r}\bigg|_{\xi=\xi_0}.
\end{equation}
Substituting into Eq. (5.3) we obtain
\begin{equation}
\label{5.14}\overline{\Omega}_2=-\frac{3}{4b^2}\int^1_{-1}\vc{e}_z\cdot(\vc{r}\times\overline{\vc{u}}_{2S})\bigg|_{\xi=\xi_0}\;d\eta,
\end{equation}
for surface velocity $\overline{\vc{u}}_{2S}$ which does not depend on the variable $\varphi$. This reduces to the result for a sphere \cite{16} in the limit $b=a$.

The identities Eqs. (4.12) and (5.13) for the surface force density in the Stokes friction problem agree with those obtained by Brenner \cite{9} and Kim \cite{10},\cite{11} for a general ellipsoid. Eq. (4.12) may be cast in the alternative form
\begin{equation}
\label{5.15}\vc{e}_1\cdot\vc{\sigma}_{Tz}\bigg|_{\xi=\xi_0}=\frac{-1}{4\pi ab^2}(\vc{e}_1\cdot\vc{r})\mathcal{F}_z\vc{e}_z\bigg|_{\xi=\xi_0},
\end{equation}
and Eq. (5.13) may be cast in the form
\begin{equation}
\label{5.16}\vc{e}_1\cdot\vc{\sigma}_{Rz}\bigg|_{\xi=\xi_0}=\frac{-3}{8\pi ab^4}(\vc{e}_1\cdot\vc{r})\mathcal{T}_z\vc{e}_z\times\vc{r}\bigg|_{\xi=\xi_0}.
\end{equation}
Kim \cite{11} has related the identities for a general ellipsoid to a symmetry property of the Stokes double-layer operator \cite{12}.

\section{\label{VI}Swimming with optimal efficiency}

Our purpose is to find the mean second order translational and rotational swimming velocities and the mean rate of dissipation for given harmonically varying surface displacement. The velocities can be evaluated from the velocity $\overline{\vc{u}}_{2S}(\vc{s})$ at the undisplaced surface by use of the simplified expressions Eqs. (4.13) and (5.14). According to Eq. (2.7) this is given by a bilinear expression in terms of the displacement $\vc{\xi}_\omega(\vc{s})$ and the first order velocity field $\vc{v}_\omega(\vc{r})$.
The mean rate of dissipation in Eq. (2.12) is given by a bilinear expression in $\vc{v}_\omega$ and $p_\omega$. By expansion of $\vc{v}_\omega(\vc{r})$ and $p_\omega(\vc{r})$ in terms of the mode functions defined in Sec. III the expressions can be written as expectation values of moment vectors incorporating the mode coefficients as components, with matrices representing the various linear operators. Subsequently the mean translational swimming velocity can be optimized for given power.

We indicate the three different types of mode functions by a discrete index $\sigma$ taking the values $(0,1,2)$. Thus we use mode functions $\{\vc{f}_{nm\sigma}(\vc{r})\}$
given by
\begin{equation}
\label{6.1}\vc{f}_{nm0}(\vc{r})=\xi^n_0\vc{v}_{nm}(\vc{r}),
\qquad\vc{f}_{nm1}(\vc{r})=\xi^{n+1}_0\vc{w}_{nm}(\vc{r}),\qquad\vc{f}_{nm2}(\vc{r})=\xi^{n+2}_0\vc{u}_{nm}(\vc{r}).
\end{equation}
The prefactors are chosen to have normalization similar to that for the mode functions of a sphere \cite{5},\cite{25}. The corresponding pressure mode functions are \begin{equation}
\label{6.2}p_{nm\sigma}(\vc{r})=\delta_{\sigma 0}\xi_0^np_{nm}(\vc{r}).
\end{equation}
The first order velocity field $\vc{v}_\omega(\vc{r})$ and pressure $p_\omega(\vc{r})$ can be expanded as
 \begin{equation}
\label{6.3}\vc{v}_\omega(\vc{r})=\omega a\sum_{nm\sigma}c_{nm\sigma}\vc{f}_{nm\sigma}(\vc{r}),\qquad p_\omega(\vc{r})=\omega a\sum_{nm\sigma}c_{nm\sigma}p_{nm\sigma}(\vc{r}),
\end{equation}
with dimensionless complex moments $\{c_{nm\sigma}\}$. The latter are regarded as components of a complex vector $\vc{\psi}$.

We consider a superposition of solutions of the form Eq. (6.3) with a single value of $m$.
Then the various bilinear products are independent of the azimuthal angle $\varphi$, and by symmetry the swimming is in the $z$ direction. The mean second order swimming velocity has value $\overline{U_2}$ given by
\begin{equation}
\label{6.4}\overline{U_2}=\frac{1}{2}\;\omega
a(\vc{\psi}|\du{B}|\vc{\psi}),
\end{equation}
with a dimensionless hermitian matrix $\du{B}$. The mean second order rotational swimming velocity is about the $z$ axis with value $\overline{\Omega_2}$ given by
\begin{equation}
\label{6.5}\overline{\Omega_2}=\frac{3}{4}\;\frac{\omega a^2}{b^2}
(\vc{\psi}|\du{C}|\vc{\psi}),
\end{equation}
with a dimensionless hermitian matrix $\du{C}$. The time-averaged rate of dissipation can be expressed as
\begin{equation}
\label{6.6}\overline{\mathcal{D}_2}=8\pi\eta_s\omega^2a^3(\vc{\psi}|\du{A}|\vc{\psi}),
\end{equation}
with a dimensionless hermitian matrix $\du{A}$. The elements of the three matrices can be evaluated from Eqs. (4.13), (5.14) and (2.12) and are determined by one-dimensional integrals.

We must put the moments $c_{1m0},c_{1m1}$ equal to zero, because of the requirement that the swimmer exert no net force or torque on the fluid. Correspondingly for $m=0,1$ the matrices $\du{A},\du{B},\du{C}$ can be truncated by deleting the first two rows and columns. In the following we assume that this truncation has been performed. We denote the matrices with additional truncation at maximum order $n$ equal to $L$ and chosen value $m$ as $\du{A}_{Lm},\;\du{B}_{Lm},\;\du{C}_{Lm}$. The matrices depend on the aspect ratio $a/b$ via $\xi_0=1/\sqrt{1-(b^2/a^2)}$, but not on the size of the spheroid.

The matrices $\du{A}_{Lm}$ and $\du{C}_{Lm}$ turn out to be real and symmetric, and the matrix $\du{B}_{Lm}$ is pure imaginary and antisymmetric. This has been achieved by the choice of phase in Eq. (3.9). As in the case of a sphere we find that the matrices take a checkerboard form with zeros on alternate positions. In the case of a sphere the matrix elements can be evaluated in analytic form \cite{5}. For a spheroid we must take recourse to numerical calculation. We have evaluated the matrices for chosen values of $\xi_0$ up to order $L=3$.

It is of interest to optimize the mean translational swimming velocity $\overline{U}_2$ for given mean rate of dissipation $\overline{\mathcal{D}}_2$. Taking the latter into account via a Lagrange multiplier $\lambda$ we are led to a generalized eigenvalue problem of the form
\begin{equation}
\label{6.7}\du{B}|\vc{\psi}_\lambda)=\lambda\du{A}|\vc{\psi}_\lambda).
\end{equation}
The normalization of the mode functions affects the matrices and the eigenvectors, but not the eigenvalues.
The maximum eigenvalue $\lambda_{max}$ determines the optimum translational swimming velocity for given power. The corresponding mean rotational swimming velocity $\overline{\Omega}_2$ can be found from the eigenvector $|\psi_\lambda)$ by use of Eq. (6.5). We denote the maximum eigenvalue with maximum order $n$ equal to $L$ and given $m$ as $\lambda_{Lm,max}$. The swimming efficiency is defined as the ratio of speed and power in the dimensionless form \cite{15}
\begin{equation}
\label{6.8}E_T=4\eta_s\omega a^2\frac{|\overline{U}_2|}{\overline{\mathcal{D}}_2}.
\end{equation}
The optimum efficiency is related to the maximum eigenvalue by
\begin{equation}
\label{6.9}E_{T,max}=\lambda_{max}/(4\pi).
\end{equation}
For a sphere the maximum eigenvalue is obtained for $m=0$ and then takes the value $2\sqrt{2}$ for maximum order $L\rightarrow\infty$, as shown earlier \cite{17}.
The maximum eigenvalue is affected by the shape of the body, but not by its size. Thus for a prolate spheroid and chosen order $L$ the maximum eigenvalue depends on the elongation $c/b$, or equivalently the aspect ratio $a/b$, but not on the value of $a$. For large elongation $c/b$ the maximum eigenvalue can be much larger than for a sphere for the same value of $L$.

In Table I we list the values of the maximum eigenvalue $\lambda_{Lm,max}$ for a sphere and for a spheroid for maximum orders $L=2$ and $L=3$ for various values of $m$. For $L=2$ and $m=0,1$ the matrices are four-dimensional. For $L=2$ and $m=2$ the matrices are three-dimensional. For $L=3$ and $m=0,1$ the matrices are seven-dimensional. For $L=3$ and $m=2$ the matrices are six-dimensional, and for $L=3,\;m=3$ they are three-dimensional. The values for a sphere are obtained for a different set of flow functions than for a spheroid. For a sphere the maximum order $L$ is the maximum polar angular number $l$ in the usual notation. By explicit comparison of the analytic expressions at order $L=2$ it can be seen that even in the limit $\xi_0\rightarrow\infty$ the two sets for given $L$ do not span the same space. The calculation for the spheroid becomes numerically difficult for large $\xi_0$.

Not surprisingly, for a spheroid the efficiency increases monotonically with increasing elongation $c/b$ in all cases shown in Table I. There is a remarkable dependence on the azimuthal number $m$. Already for small elongation the swimming for $L=2$ at $m=1$ is more efficient than at $m=0$, but note that for $L=3$ and $c/b=6$ the swimming at $m=0$ is slightly more efficient than at $m=1$, and the swimming at $m=2$ is more efficient than at $m=0,1$.

In Fig. 1 we plot the maximum eigenvalue $\lambda_{Lm,max}$ as a function of elongation $c/b$ for $L=2$ and $m=(0,1,2)$. This shows that for large elongation the swimming for $m=1$ is the most efficient. In Fig. 2 we show similar plots for $L=3$ and $m=(0,1,2,3)$. Again for large elongation the swimming for $m=1$ is the most efficient. We compare the curve for $m=1$ with asymptotic behavior of the form $\log_{10}[1.611/(\varepsilon\log\varepsilon)^2)]$, where $\varepsilon=b/a=1/\sqrt{c^2/b^2+1}$, with coefficient $1.611$ chosen to fit the numerical value at $c/b=18$. The form is suggested by that found by Leshansky et al. \cite{6}.

In Fig. 3 we show the displacement in the normal direction in the plane $\varphi=0$ for optimal swimming of a spheroid of elongation $c/b=18$ with modes up to $L=3$ with $m=1$ at times $t_j=jT/16$, where $T=2\pi/\omega$ is the period and $j=0,1,...,8$. The normal component has a simple dependence on the polar variable $\eta$, concentrating the displacement at both ends of the spheroid. In the azimuthal direction there is a running wave given by the factor $\exp[i(\varphi-\omega t)]$.

For $m\neq 0$ the rotational swimming velocity does not vanish for the eigenvector $\vc{\xi}_{Lm}$ corresponding to $\lambda_{Lm,max}$. In Fig. 4 we plot the ratio
\begin{equation}
\label{6.10}\rho_{Lm}=\frac{(\vc{\xi}_{Lm}|\du{C}_{Lm}|\vc{\xi}_{Lm})}{(\vc{\xi}_{Lm}|\du{A}_{Lm}|\vc{\xi}_{Lm})}
\end{equation}
as a function of elongation $c/b$ for $L=2$ and $m=(1,2)$. For $m=0$ the ratio vanishes. In Fig. 5 we show similar plots for $L=3$ and $m=(1,2,3)$.

We denote the reduced power of the optimal swimmer with moments $\vc{\xi}_{Lm}$ as $P_{Lm}$,
\begin{equation}
\label{6.11}P_{Lm}=(\vc{\xi}_{Lm}|\du{A}_{Lm}|\vc{\xi}_{Lm}).
\end{equation}
The time the swimmer needs to move over a distance equal to the length of the spheroid is \cite{27}
\begin{equation}
\label{6.12}t_{Lm}=\frac{2a}{\overline{U}_2}=\frac{4}{\omega\lambda_{Lm}P_{Lm}}.
\end{equation}
During this time the swimmer rotates over  the angle
\begin{equation}
\label{6.13}\overline{\Omega}_2t_{Lm}=\frac{3a^2|\rho_{Lm}|}{b^2\lambda_{Lm}}.
\end{equation}
This is independent of the power. For a spheroid with elongation $c/b=18$ and modes up to $L=3$ with $m=1$ the right hand side takes the value $11.525$.

\section{\label{VII}Discussion}

We have studied the optimal mode of swimming of a prolate spheroid for several classes of modes of low multipole order. It is shown in Figs. 1 and 2 that for large aspect ratio optimal efficiency is obtained for azimuthal number $m=1$, corresponding to a non-vanishing mean rotational swimming velocity. Although the calculations were performed to second order in the amplitude of surface deformation, it may be expected that the results are indicative also for large amplitude swimming at low Reynolds number.

The calculations are less complete than for a sphere \cite{5}, since for a spheroid the elements of the three crucial matrices of the bilinear theory cannot be evaluated in analytic form. However, for any given aspect ratio the matrices can be calculated numerically once and for all.

The important identities in Eqs. (5.15) and (5.16) show that the theory can be extended to a general ellipsoid. The low order mode functions for an ellipsoid are known in explicit form \cite{28}.

Finally, the extension of the bilinear theory for a sphere to a fluid with inertia \cite{29} suggests that such an extension is also possible for a prolate spheroid. It would be of interest to relate such a theory to the classic work of Lighthill on slender body fish locomotion \cite{30}.

\newpage
\section*{Table caption}
Table of values of the maximum eigenvalue $\lambda_{Lm,max}$ for various values of the elongation $c/b$ or aspect ratio $a/b$. In the first column we list corresponding values for a sphere.
\newpage
\newpage
\begin{table}[!htb]  \footnotesize\centering
  \caption{}\label{tab:1}
\begin{tabular}{|c|c|c|c|c|c|c|c|c|c|}
\hline
 $c/b$&$0$  &$0.2$& $1$ &$3$ &$6$&$9$ &$12$&$15$&$18$
 \rule[-5pt]{0pt}{16pt} \\
  $a/b$&$1$  &$1.020$& $1.414$ &$3.162$ &$6.083$&$9.055$ &$12.042$&$15.033$&$18.028$
 \rule[-5pt]{0pt}{16pt} \\
\hline
 $L=2,m=0$&$1.179$ &$1.195$& $1.317$ &$2.618$ &$5.064$&$7.049$ &$8.004$&$8.711$&$9.263$
 \rule[-5pt]{0pt}{16pt} \\
 $L=2,m=1$&$1.085$ &$1.343$& $2.056$ &$5.222$ &$10.154$&$15.252$ &$20.430$&$25.634$&$30.846$
 \rule[-5pt]{0pt}{16pt} \\
 $L=2,m=2$&$1.118$ &$1.162$& $1.930$ &$2.749$ &$2.992$&$3.090$ &$3.140$&$3.169$&$3.188$
 \rule[-5pt]{0pt}{16pt} \\
 $L=3,m=0$&$1.585$ &$1.567$& $2.326$ &$6.241$ &$13.526$&$21.288$ &$29.255$&$37.341$&$45.509$
 \rule[-5pt]{0pt}{16pt} \\
 $L=3,m=1$&$1.573$ &$1.606$& $2.443$ &$6.431$ &$13.439$&$23.345$ &$35.270$&$48.454$&$62.590$
 \rule[-5pt]{0pt}{16pt} \\
 $L=3,m=2$&$1.577$ &$1.549$& $2.434$ &$6.169$ &$14.786$&$24.428$ &$34.276$&$44.162$&$54.041$
 \rule[-5pt]{0pt}{16pt} \\
 $L=3,m=3$&$1.296$ &$1.344$& $2.162$ &$2.971$ &$3.072$&$3.096$ &$3.106$&$3.111$&$3.113$
 \rule[-5pt]{0pt}{16pt} \\
\hline
\end{tabular}
\end{table}
\newpage

\section*{Figure captions}

\subsection*{Fig. 1}
Plot of the maximum eigenvalue $\lambda_{2m,max}$ as a function of elongation $c/b$ for $m=0$ (drawn), $m=1$ (long dashes), $m=2$ (short dashes). This characterizes the mean translational swimming velocity
of the optimal swimmer for $L=2$ at each value of $m$ for given power.

\subsection*{Fig. 2}
Plot of the maximum eigenvalue $\lambda_{3m,max}$ as a function of elongation $c/b$ for $m=0$ (drawn), $m=1$ (middle dashes), $m=2$ (short dashes), and $m=3$ (long dashes). The thick drawn curve is an asymptotic approximation explained in Sec. VI.

\subsection*{Fig. 3}
Plot of the real part of the normal displacement $\xi_1$ in the meridional plane $\varphi=0$ at times $t_j=jT/16$, where $T=2\pi/\omega$ is the period and $j=0,1,...,8$, as a function of polar variable $\eta$ for the optimal swimmer with $L=3,\;m=1$ for elongation $c/b=18$.

\subsection*{Fig. 4}
Plot of the ratio $|\rho_{2m}|$, defined in Eq. (6.10), for $m=1$ (drawn), $m=2$ (dashed), as a function of elongation $c/b$. This characterizes the rate
of steady rotation of the optimal swimmer at each value of $m$ for given power.

\subsection*{Fig. 5}
Plot of the ratio $|\rho_{3m}|$, defined in Eq. (6.10), for $m=1$ (drawn), $m=2$ (long dashes), $m=3$ (short dashes), as a function of elongation $c/b$.

\newpage
\clearpage
\newpage
\setlength{\unitlength}{1cm}
\begin{figure}
 \includegraphics{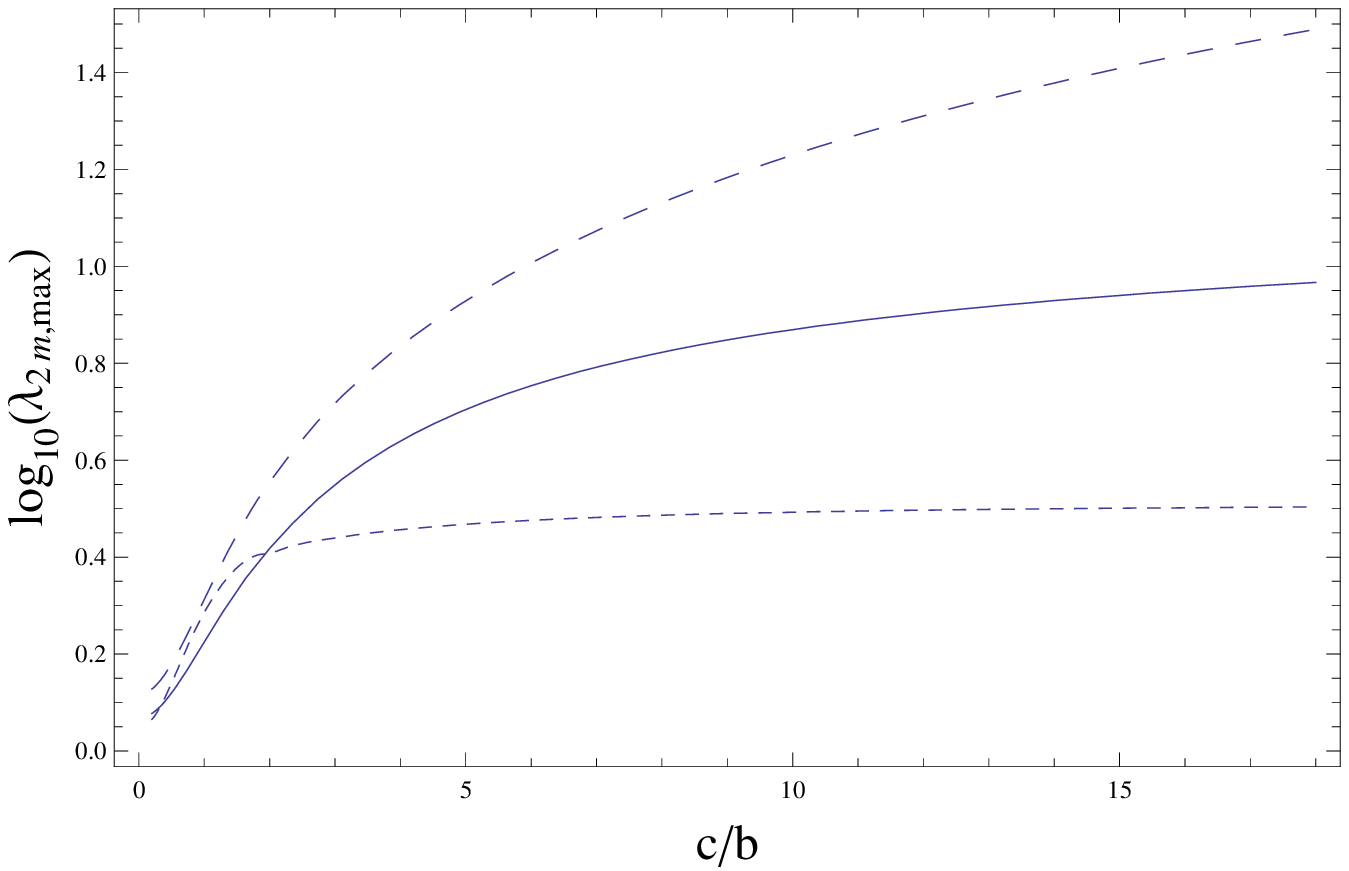}
   \put(-9.1,3.1){}
\put(-1.2,-.2){}
  \caption{}
\end{figure}
\newpage
\newpage
\clearpage
\newpage
\setlength{\unitlength}{1cm}
\begin{figure}
 \includegraphics{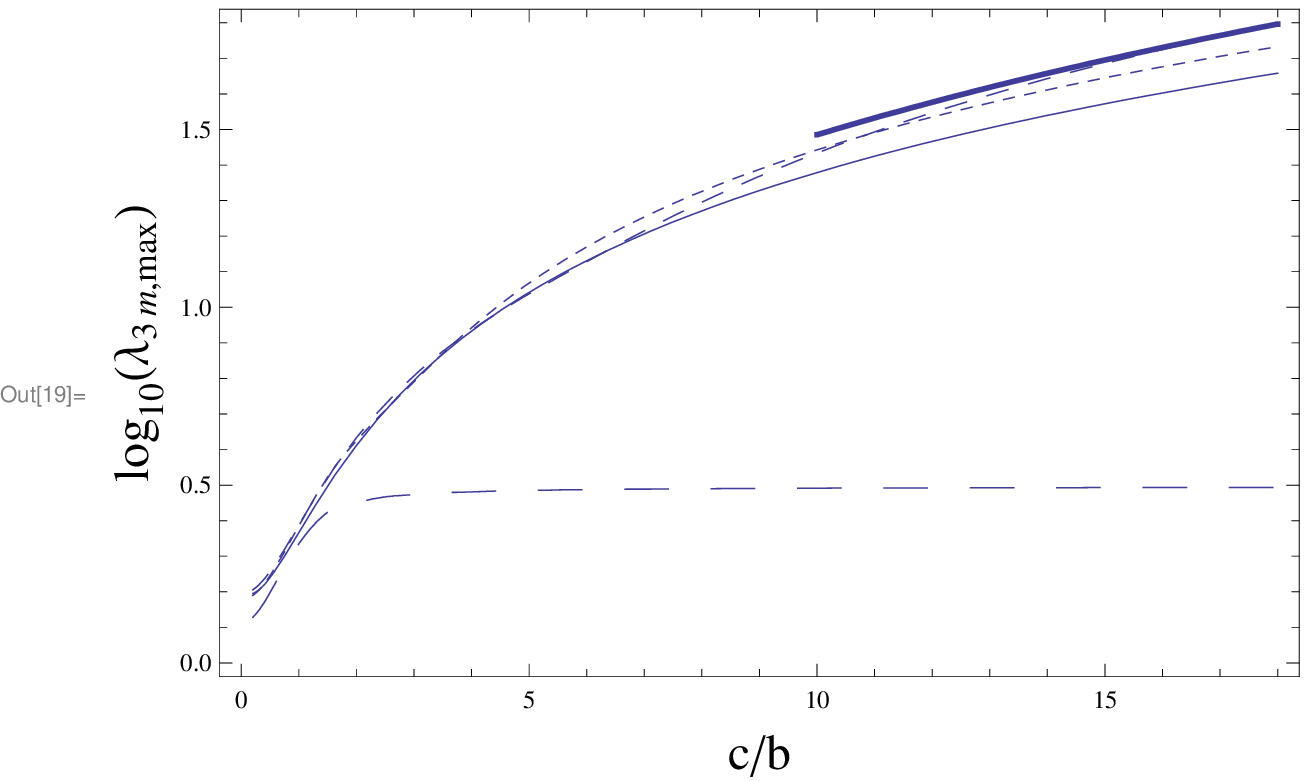}
   \put(-9.1,3.1){}
\put(-1.2,-.2){}
  \caption{}
\end{figure}
\newpage
\newpage
\clearpage
\newpage
\setlength{\unitlength}{1cm}
\begin{figure}
 \includegraphics{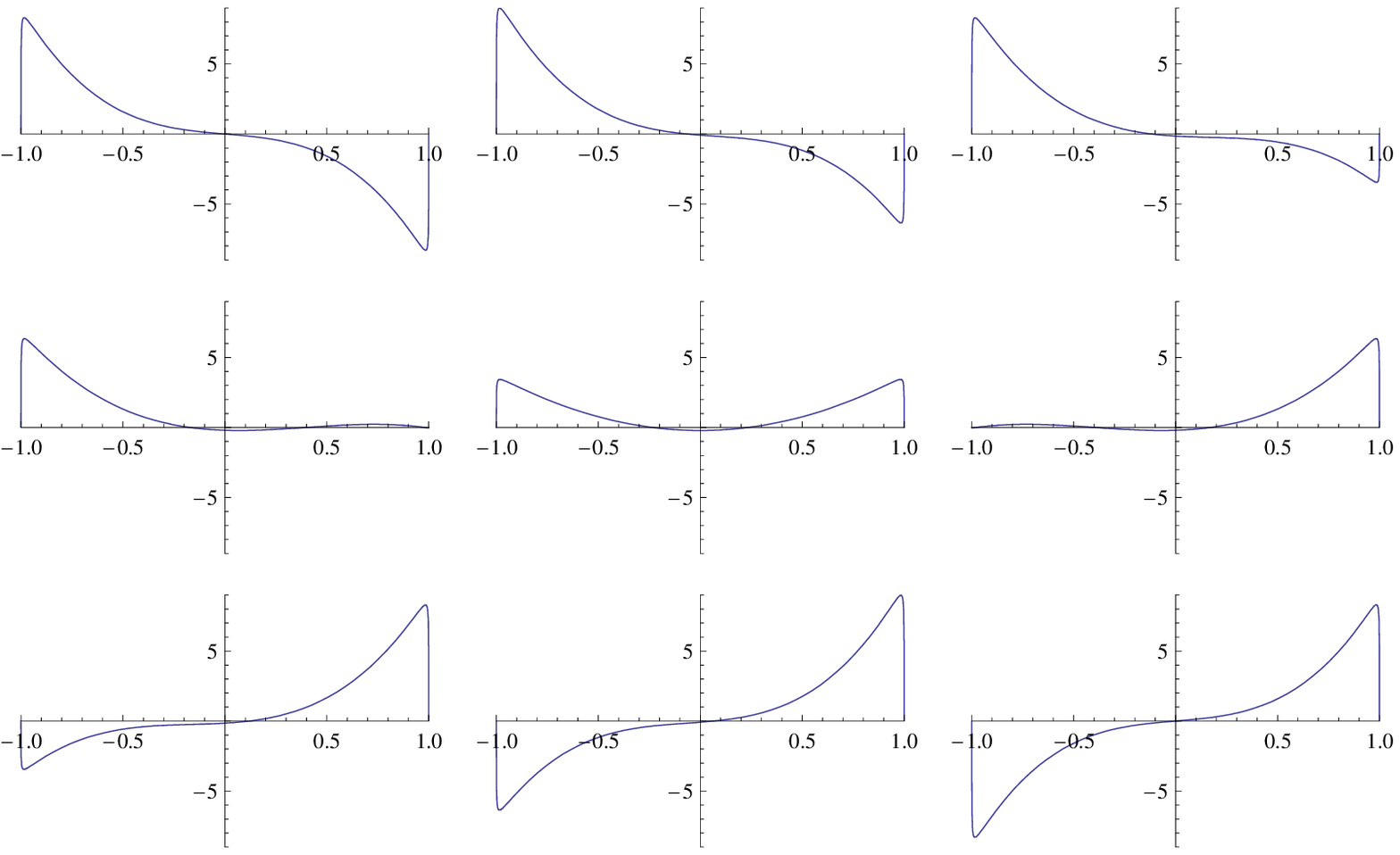}
   \put(-9.1,3.1){}
\put(-1.2,-.2){}
  \caption{}
\end{figure}
\newpage
\newpage
\clearpage
\newpage
\setlength{\unitlength}{1cm}
\begin{figure}
 \includegraphics{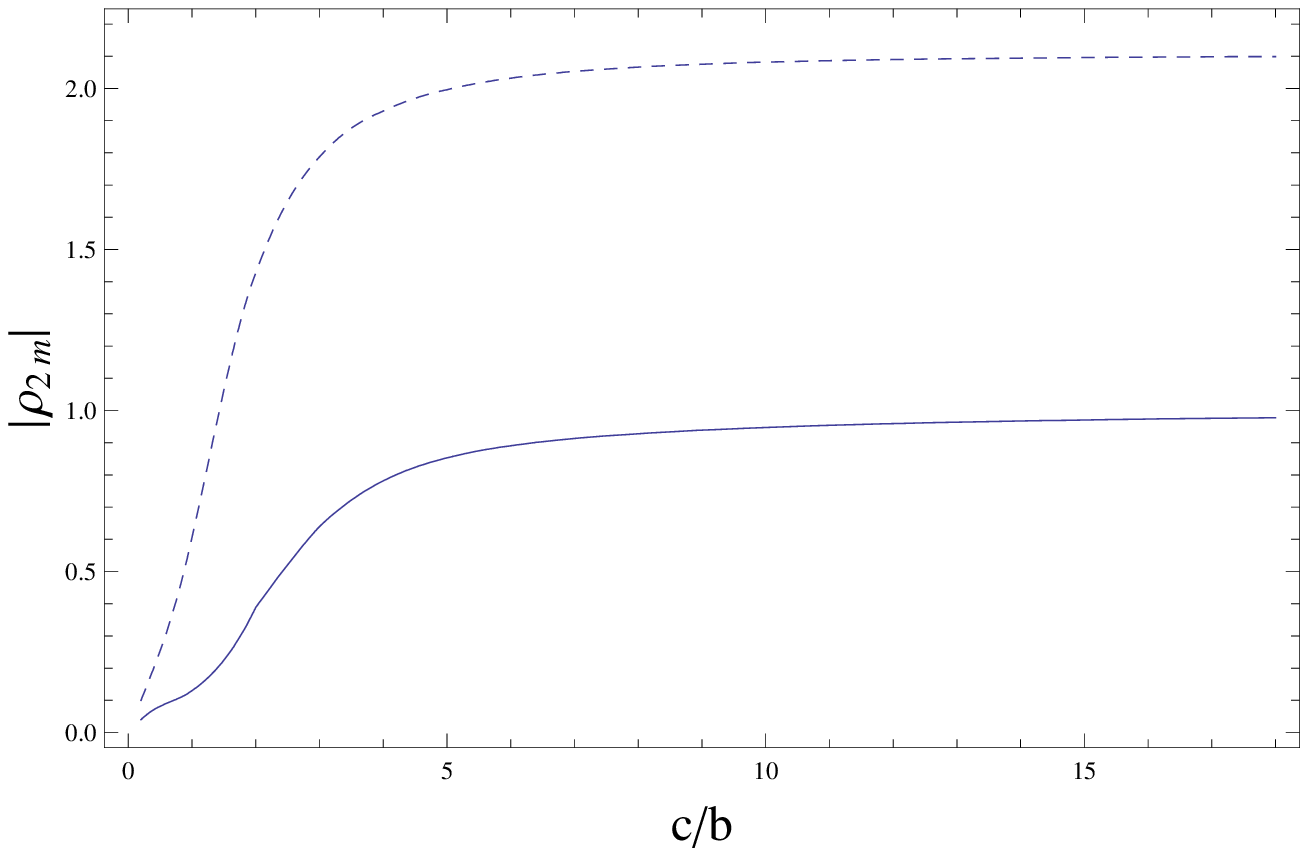}
   \put(-9.1,3.1){}
\put(-1.2,-.2){}
  \caption{}
\end{figure}
\newpage
\clearpage
\newpage
\setlength{\unitlength}{1cm}
\begin{figure}
 \includegraphics{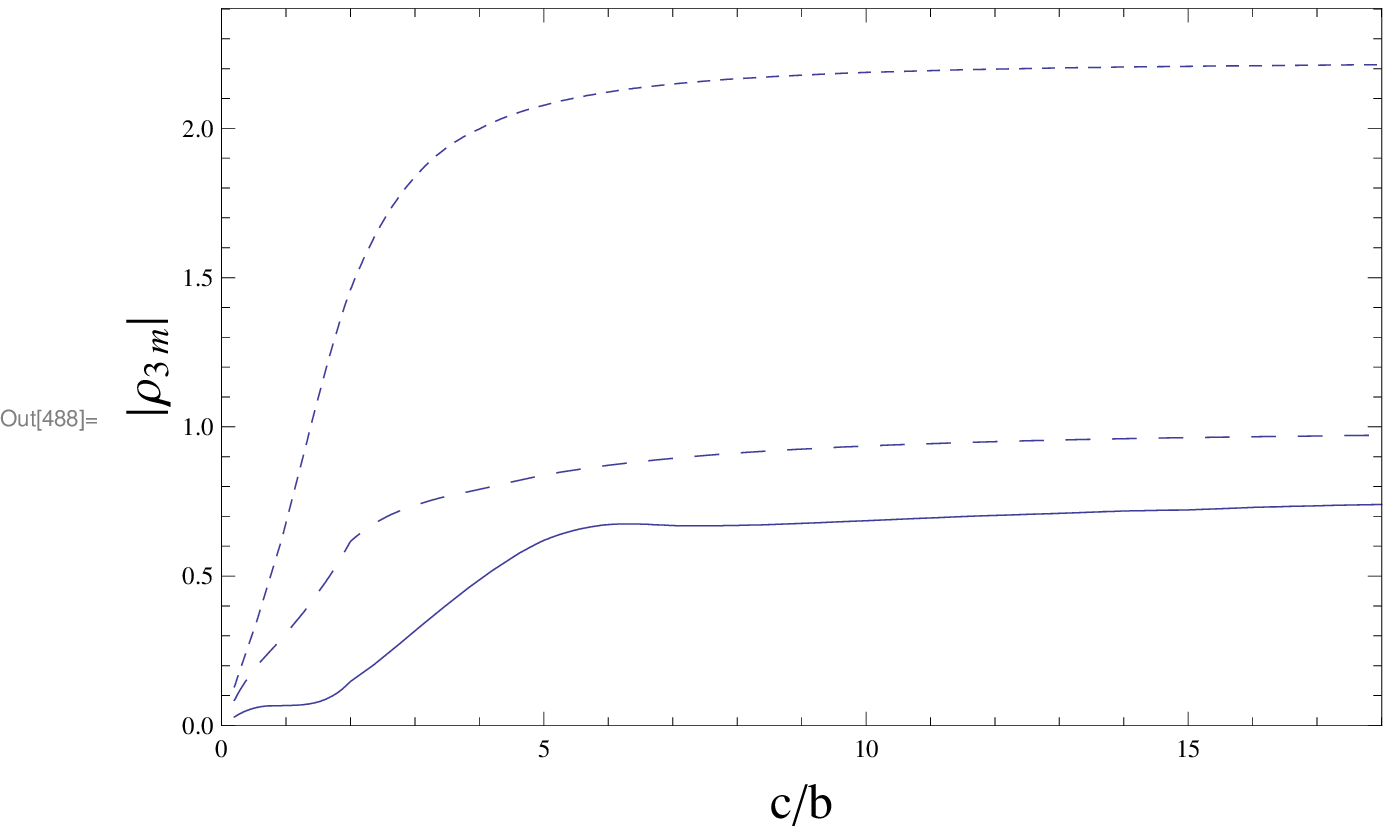}
   \put(-9.1,3.1){}
\put(-1.2,-.2){}
  \caption{}
\end{figure}
\newpage
\end{document}